\documentclass[twocolumn,showpacs,superscriptaddress,amsmath,amssymb,prl]{revtex4}

\usepackage{graphicx}%
\usepackage{dcolumn}%

\begin{document}

\title{From tunneling to photoemission: correlating two spaces}

\author{A. A. Kordyuk}
\affiliation{Institute of Metal Physics of National Academy of Sciences of Ukraine, 03142 Kyiv, Ukraine}
\affiliation{Leibniz-Institut f\"ur Festk\"orper- und Werkstoffforschung Dresden, 01069 Dresden, Germany}

\author{V. B. Zabolotnyy}
\author{D. S. Inosov}
\author{S. V. Borisenko}
\affiliation{Leibniz-Institut f\"ur Festk\"orper- und Werkstoffforschung Dresden, 01069 Dresden, Germany}

\date{November 1, 2005}%

\begin{abstract}
Correlating the data measured by tunneling and photoemission spectroscopies is a long-standing problem in condensed matter physics. The quasiparticle interference, recently discovered in high-$T_c$ cuprates, reveals a possibility to solve this problem. Application of modern phase retrieval algorithms to Fourier transformed tunneling data allows to recover the distribution of the quasiparticle spectral weight in the reciprocal space of solids measured directly by photoemission. This opens a direct way to unify these two powerful techniques and may help to solve a number of problems related with space/time inhomogeneities predicted in strongly correlated electron systems.
\end{abstract}


\preprint{\textit{xxx}}

\maketitle

The electronic inhomogeneity in high-$T_c$ cuprates (HTSC) seen by scanning tunneling spectroscopy (STS) \cite{1} has attracted much interest from the scientific community because of its clear relation to a central problem of HTSC-the mechanism of their evolution with hole doping from the antiferromagnetic insulator to superconductor. Recent breakthrough in the development of STS technique to a level where the Fourier transformed (FT) STS images have been crystallized into well-defined symmetric patterns \cite{2} has revealed the existence of regular inhomogeneities which could relate the HTSC problem to self-ordering phenomena in correlated electron systems. An important step in this direction have been made by identifying of the essential part of these inhomogeneities with rather trivial quasiparticle interference (QI) through an impurity scattering hypothesis \cite{2,3,3a}, which says that the FT STS function, $\mathbf{F} S(\mathbf{r})$, is proportional to a joint density of states $C(\mathbf{q})$ which, in turn, is just an autocorrelation (AC) of the quasiparticle spectral function $A(\mathbf{k})$. However, a reason for the remaining non-QI inhomogeneities, as well as the scattering hypothesis by itself, have yet to be understood. This is due, in part, to lack of deep understanding of the scattering phenomenon, but mainly due to an absence of a direct transition from the joint $\mathbf{q}$-space of FT STS to the reciprocal $\mathbf{k}$-space where the distribution of the quasiparticle spectral function, $A(\mathbf{k})$, is known from the angle resolved photoemission spectroscopy (ARPES) with a tremendous accuracy \cite{4}. The AC procedure gives an attractive possibility to compare the FT STS images to the AC ARPES data but evidently suffers from the absence of a direct transformation from the joint to plain reciprocal spaces: Due to presence of experimental factors such as matrix elements the process of fitting $A(\mathbf{k})$ to $C(\mathbf{q})$ is time consuming and, more important, does not give a unique solution \cite{STSvsARPES, STSvsARPESb}. Here we suggest a procedure to uniquely recover $A(\mathbf{k})$ from $S(\mathbf{r})$, shifting the problem of STS and ARPES correlation to ARPES domain.

\begin{figure*}[!t]
\includegraphics[width=16cm]{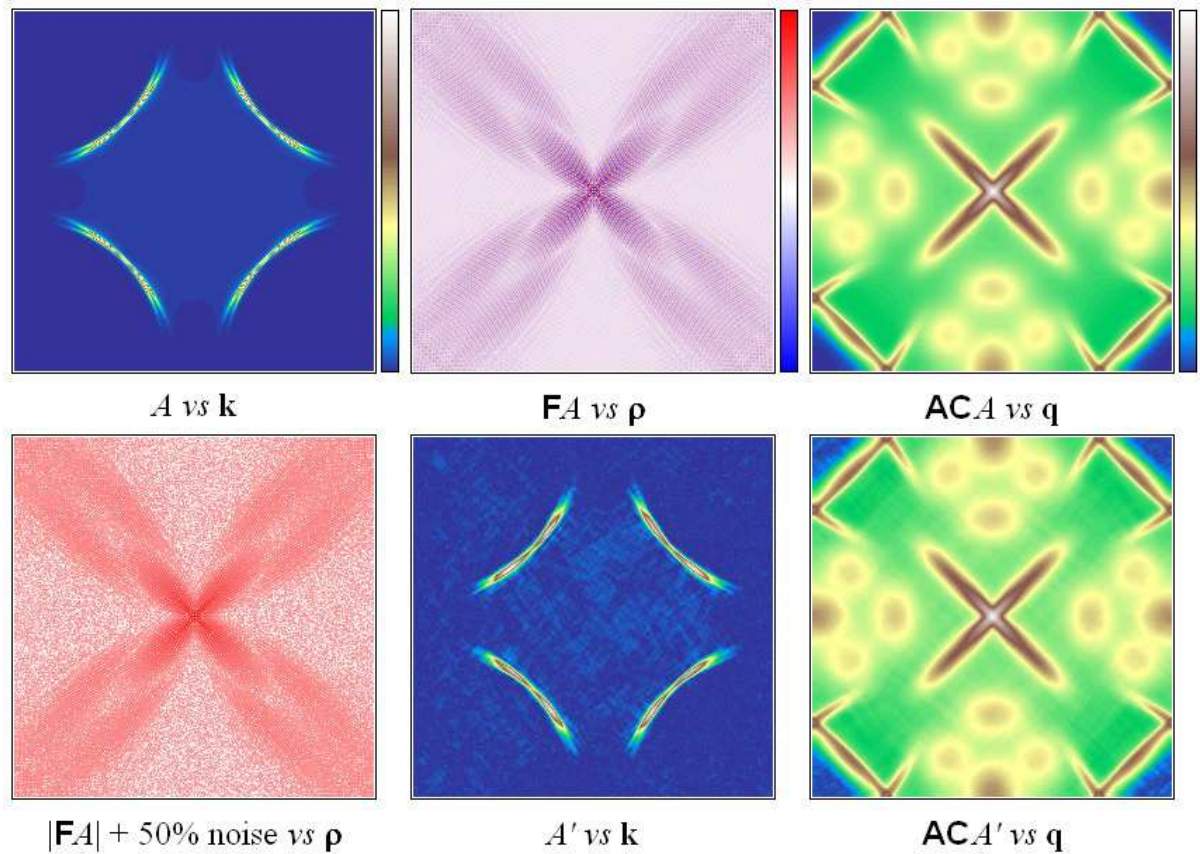}%
\caption{\label{Images} Representations of quasiparticles in different spaces of high-$T_c$ cuprates: the spectral function $A$ in the reciprocal space $\mathbf{k}$, the real part of its Fourier image $\mathbf{F} A$ in a joint real space $\mathbf{\rho}$, and a joint spectral weight $C = \mathbf{AC}\;A$ in a joint reciprocal space $\mathbf{q}$. The reverse transformation $C \rightarrow A$ is possible by means of PRA (see Movie S1). The bottom row illustrates the stability of PRA to experimental uncertainty---a noise has been added to $|\mathbf{F} A|$ (see also Movie S2). $A(\mathbf{k})$ is simulated at 20 meV binding energy based on experimentally determined parameters of the bare band dispersion for an optimally doped Bi-2212 compound \cite{6} taking into account a $d$-wave superconducting gap $\Delta = \Delta_0 (\cos k_x - \cos k_y)/2$ with $\Delta_0$ = 40 meV; $\mathbf{k}$- and $\mathbf{q}$-images cover exactly the 1st Brillouin zone.}
\end{figure*}

To understand the essence of the concept, let us rewrite the above mentioned impurity scattering hypothesis in a symbolic way: 
\begin{equation}\label{ISH}
|\mathbf{F} S| = \mathbf{AC}\; A = C.
\end{equation}
The AC is defined as
\begin{equation}\label{AC}
\mathbf{AC}\; A(\mathbf{k}) = \int{A(\mathbf{k}) A(\mathbf{k} + \mathbf{q}) d\mathbf{k}} = C(\mathbf{q}),
\end{equation}
and is therefore an irreversible and time consuming operation. 

Nevertheless, we have found that both problems can be solved using the Wiener-Khinchin theorem \cite{WK}, which sets an extremely important relationship between AC and FT: 
\begin{equation}\label{WKT}
\mathbf{AC}\; A = \mathbf{F}^{-1} |\mathbf{F} A|^2.
\end{equation}
The fast Fourier routine essentially speeds up the calculations, but more important is that the theorem shows the way to reverse the autocorrelation procedure and derive $A(\mathbf{k})$ from $S(\mathbf{r})$: With Eq.~(\ref{WKT}), the impurity scattering hypothesis can be rewritten as
\begin{equation}\label{MyF}
|\mathbf{F} A| = \sqrt{\mathbf{F}|\mathbf{F} S|},
\end{equation}
and the problem is reduced to recovering of a function from its Fourier amplitude $R(\mathbf{\rho}) = |\mathcal{R(\mathbf{\rho})}| = |\mathbf{F} A(\mathbf{k})|$. 

Fortunately, similar problem arose in applied optics and a number of phase retrieval algorithms (PRA) were developed \cite{5}. These algorithms involve iterative Fourier transformation back and forth between the object and Fourier domains with application of the known constraints.

Fig.~1 (top row) shows $A(\mathbf{k})$---a gaped quasiparticle density of states of an optimally doped CuO-bilayer---as well as the real part of its Fourier image, $\mathrm{Re}(\mathbf{F}A)$, and the result of its autocorrelation, $\mathbf{AC}\;A$. $A(\mathbf{k})$ is simulated at 20 meV binding energy based on experimentally determined parameters of the bare band dispersion for an optimally doped Bi-2212 compound \cite{6} taking into account a $d$-wave superconducting gap $\Delta = \Delta_0 (\cos k_x - \cos k_y)/2$ with $\Delta_0$ = 40 meV. We note, that $\mathrm{Im}(\mathbf{F}A)$ = 0 due to even symmetry of the spectral function, $A(\mathbf{k})=A(\mathbf{-k})$, so, the problem can be reduced to recovering of a \textit{real} smooth function from its modulus. It seems that such a procedure can be realized, in principle, but should evidently suffer from finite resolution and unavoidable uncertainty of the experiment. 

In turn, the PRA are essentially discrete. Their uniqueness, in case of positive and spatially confined object, has been proved theoretically while the stability to noise and speed of convergence are the issues of continues development \cite{5}. The recovering of $A(\mathbf{k})$ from $|\mathbf{F} A|$ by means of PRA is illustrated by Movie 1 \cite{S}. Here we used a modified ``input-output" algorithm \cite{5}, the $n$th iteration of which can be formulated as follows:
\begin{eqnarray}
\mathcal{R}_{n} &=& \mathbf{F} A_{n},\\
\mathcal{R}'_{n} &=& R \exp[i \arg(\mathcal{R}_{n})],\\
\mathcal{A}'_{n} &=& \mathbf{F}^{-1}\mathcal{R}'_{n},
\end{eqnarray}
\begin{eqnarray}
A_{n+1} &=& 
    \begin{cases}
    \mathrm{Re}(\mathcal{A}'_{n}) & \text{if $\mathrm{Re}(\mathcal{A}'_{n}) \ge 0$}, \\
    \mathrm{Re}(A_{n}-\beta \mathcal{A}'_{n}) & \text{if $\mathrm{Re}(\mathcal{A}'_{n}) < 0$},
    \end{cases}
\end{eqnarray}
where $R(\mathbf{\rho})$ is a ``source'' function, and $\beta$ is a constant which we choose between 1 and 2 as a compromise between speed of convergence and stability of the algorithm. As an initial guess we used a Gaussian distribution with random noise: $A_0(\mathbf{k}) = \exp(|\mathbf{k}|^2/w^2) + noise$.

Bottom row of Fig.~1 illustrates the robustness of PRA to experimental uncertainty: $A'(\mathbf{k})$ is recovered from a noisy $|\mathbf{F} A|$ (the noise has been simulated by a random value from a Gaussian distribution such that the standard deviation of an infinite number of such values would be 50\% of the average value of $|\mathbf{F} A|$). The work of RPA in this case is illustrated by Movie 2 \cite{S}.

Finally, we discuss the existent attempts to compare the STS and ARPES data in the $\mathbf{q}$-domain \cite{3, STSvsARPES, STSvsARPESb}. It has been shown that the intensity maps measured by ARPES, when autocorrelated, does not result in so distinct spots as observed in FT STS images \cite{3}. The correspondence can be made better assuming better energy resolution in ARPES \cite{3}, or $\mathbf{k}$-dependent matrix elements in STS \cite{STSvsARPES}.  We believe that both effects should be taken into account together with the gaped and highly anisotropic quasiparticle self-energy, but, in this paper, we do not purpose to define a region of parameters which result in spot patterns seen in FT STS. Instead, we suggest the approach to recover $A(\mathbf{k})$ from FT STS shifting this problem to the ARPES domain. Comparing $A(\mathbf{k})$ directly measured by ARPES to one determined from STS with the suggested approach will, in particular, clarify such long standing problems as tunneling matrix elements \cite{Bart} and inconsistency in the values of the quasiparticle self-energies determined from STS and ARPES experiments on high-$T_c$ cuprates \cite{Bart}.

In summary, we have proposed a procedure to derive the distribution of the quasipartical spectral weight in plain reciprocal space of a crystal from STS measurements and demonstrated its robustness to experimental uncertainty. When applied to STS data, this procedure will give an ultimate check for the validity of the impurity scattering hypothesis and, if valid, open a direct way to unify STS and ARPES techniques.

The project is part of the Forschergruppe FOR538.

\end{document}